# Reachability Under Uncertainty

**Allen Chang and Eyal Amir**
Department of Computer Science
University of Illinois at Urbana-Champaign
Urbana, IL 61801
{achang, eyal}@uiuc.edu

## Abstract

In this paper we introduce a new network reachability problem where the goal is to find the most reliable path between two nodes in a network, represented as a directed acyclic graph. Individual edges within this network may fail according to certain probabilities, and these failure probabilities may depend on the values of one or more hidden variables. This problem may be viewed as a generalization of shortest-path problems for finding minimum cost paths or Viterbi-type problems for finding highest-probability sequences of states, where the addition of the hidden variables introduces correlations that are not handled by previous algorithms. We give theoretical results characterizing this problem including an NP-hardness proof. We also give an exact algorithm and a more efficient approximation algorithm for this problem.

## 1 INTRODUCTION

In this paper we study a new type of inference problem where the goal is to find the most reliable path from a given source to a given sink through a network where edges within the network may fail according to certain probabilities. Finding minimum cost or most reliable paths in a network is a classic computer science problem; in this paper, we consider a novel but natural variant where edge failure probabilities may depend on the state of a hidden random variable. The introduction of hidden state adds complexities to the problem of finding most reliable paths not present in previous variants; in particular, hidden state can be used to model correlations between edge failures that cannot be modeled by previous variants. In this paper, we give a number of theoretical results characterizing the presented problem including an NP-hardness proof which shows that computing an exact solution is impractical in the general case. We also give an exact algorithm and a faster approximation algorithm for solving the presented problem.

We believe algorithms which tackle this problem will find useful applications in computer network QoS (quality-of-service) routing. Existing techniques for QoS routing generally do not model any hidden state of the network, and furthermore, edge failures are assumed to be independent [Orda, 1999, F. A. Kuipers and Mieghem, 2002]. By considering hidden states as well as allowing for correlations between failures, we believe our problem formulation can be used to produce more realistic models of networks which may lead to improved routing performance. For example, hidden state variables could be used to model network hardware or software, where failures between nodes or links running on similar systems are often correlated (as they may share vulnerabilities).

We also believe algorithms for tackling the problem we introduce will find useful applications involving weighted finite state automata. Weighted finite state automata are currently used in pattern recognition and parsing applications [Smith and Eisner, 2005, Thollard et al., 2005a, Thollard et al., 2005b], where inference procedures typically involve finding most-likely sequences of states or parses of a given automaton. In prior applications utilizing weighted finite state automata, it has been assumed that transitions between states are not correlated and that there is no hidden state. Again, by introducing hidden variables, we believe richer models may be built which may lead to improved performance in applications using these automata.

There has been a substantial amount of theoretical algorithms and operations research that has tackled problems similar in flavor to the problem we propose in this paper. This research includes work on restricted shortest-path problems [Ergun et al., 2002, Hassin, 1992, Lorenz and Raz, 1999], bicriteria net-work design problems [Marathe et al., 1998], and other similar shortest-path problem formulations [Nikolova et al., 2006, Loui, 1983]. This previous work considers cases where there are multiple objectives or where the objective to is to minimize some non-standard criterion. Also worth mentioning is the work on percolation theory, which describes connectivity within graphs where edges within the graph fail stochastically [Grimmett, 1989]. Percolation theory, unlike our work, focuses on asymptotic behavior of highly-regular infinite graphs (e.g. the square lattice). To our knowledge, none of this previous work addresses or can be used to address the specific issue of hidden state which we introduce.

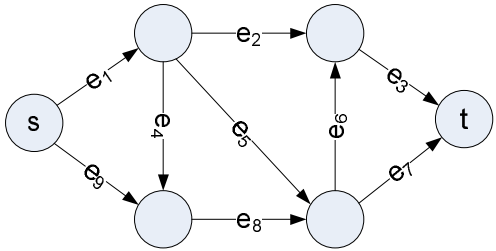

Figure 1: An example network.

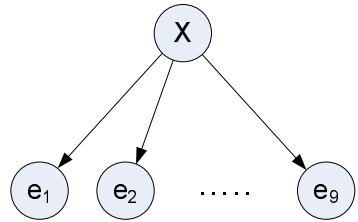

Figure 2: The graphical model describing the probability distribution for the random variables in the example network.

## 2  DEFINITIONS

In this section we will formally describe the problem that we are trying to solve. Let $G = (V, E)$ describe a directed acyclic graph, and let $s, t \in V$ denote two vertices of $G$. Let $s$ be called the source vertex, and let $t$ be called the sink vertex. An example network is shown in Figure 1. Every edge $e \in E$ will correspond to a binary-valued random variable denoting whether the corresponding edge in the graph has "failed." That is, the event $\{e = 1\}$ corresponds to the event that the corresponding edge in the network has failed, and likewise, the event $\{e = 0\}$ corresponds to the event that the corresponding edge has not failed. Additionally, let us define a hidden random variable $\mathcal{X}$ which can take on discrete values from a finite set $\mathcal{V} = \{1, 2, \ldots, d\}$.

We will assume that we are given a probability distribution for the random variables. Furthermore, we will assume that the probability any given edge $e \in E$ fails may depend only on the value of a hidden variable $\mathcal{X}$. In other words, using graphical models terminology, we will assume that the joint probability distribution over all the hidden variables factorizes as follows:

$$\Pr[e_1 = a_1, e_2 = a_2, \cdots, e_m = a_m, \mathcal{X} = j]$$
$$= \Pr[\mathcal{X} = j] \prod_{i=1}^{m} \Pr[e_i = a_i | \mathcal{X} = j] \quad (1)$$

where $e_1, \ldots, e_m$ denote all the edges in $E$. Figure 2 shows the graphical model describing the probability distribution for the random variables. We will say that a path $\pi$ in $G$ has failed if and only if at least one edge on the path has failed. In other words, if the path $\pi$ consists of the edges $e_1, \ldots, e_k$, we have that the probability $\Pr[\pi \text{ has not failed}]$ is equal to the marginal probability $\Pr[e_1 = 0, \cdots, e_k = 0]$.

The problem we pose now is to find the *most reliable path* from vertex $s$ to vertex $t$. That is, given the graph $G$ and the probability distribution (1) we wish to find

$$\arg \max_{\pi \in \Pi} \Pr[\pi \text{ has not failed}]$$

where $\Pi$ denotes the set of all paths in $G$ from $s$ to $t$.

It is worth noting that in the trivial case that the hidden variable $\mathcal{X}$ can take on only one possible value (i.e., the hidden variable is irrelevant), then the most reliable path can be easily found in polynomial time. In such a case, all the random variables corresponding to edge failures for each edge can be taken to be independent of one another (because the value of the hidden variable must be fixed). Therefore, for instance, we may assign an additive cost to each edge where the cost of vertex $e$ is defined to be $\log(1/\Pr[e \text{ failed}])$. Thus computing the most reliable path is equivalent to finding the minimum cost path, which can be computed using an algorithm such as Dijkstra's.

On the other hand, finding the most reliable path is much more difficult in the general case where the hidden variable can take on multiple values. In such cases, the hidden variable introduces dependencies between edge failures. In particular, we will show in a later section that finding the most reliable path in the general case is NP-hard.

# 3 RESULTS

In this section, we give a number of theoretical results which characterize the problem of finding the most reliable path with hidden state.

## 3.1 HARDNESS

In this subsection, we present the following theorem which proves that finding an exact solution to the problem is intractable in general. We will consider the (easier) decision version of the optimization problem of finding the most reliable path: given a network and a probability $p$, is there a path from the source to sink that will not fail with probability at least $p$? We note that the decision version of the problem is in NP—given a candidate solution (a path from the source to sink), it is simple to compute the path's probability of failure in polynomial time. It remains to show that the problem is NP-hard.

**Theorem 3.1.** *The problem of finding the most reliable path with hidden state is NP-complete.*

*Proof.* We will consider a two-step reduction. First we will reduce a "string template matching" problem to the problem we have described. Then we will show that the string template matching problem is NP-hard. The problem is the following:

> String template matching:
>
> **Input:** A set $S$ of $d$ binary string templates $s^{(1)}, \ldots, s^{(d)} \in \{0, 1, *\}^n$. The '$*$' symbol corresponds to a wildcard. A binary string $s \in \{0, 1\}^n$ matches a binary string template $s^{(i)}$ if all non-wildcard symbols in $s^{(i)}$ match the corresponding symbols in $s$.
> **Query:** Is there a binary string that matches $\geq k$ templates in $S$?

The reduction from string template matching to the problem of finding the most reliable path can be seen as follows. The hidden variable $\mathcal{X}$ will be allowed to take on the values $1, \ldots, d$. The prior probability distribution of $\mathcal{X}$ will be uniformly distributed. The structure of the graph will be that there are $n+1$ nodes numbered $1, \ldots, n+1$. Between each pair of nodes $i$ and $i+1$, there will be two edges—call these $e_i$ and $e'_i$. (We may choose to disallow multi-edges, the extension is trivial.) The conditional probability distributions of edge failures will be defined as follows. We define $\Pr[e_i \text{ fails} | \mathcal{X} = j] = 0$ if $s_i^{(j)} = 1$ and 1 otherwise. Likewise, define $\Pr[e'_i \text{ fails} | \mathcal{X} = j] = 0$ if $s_i^{(j)} = 0$ and 1 otherwise. By our construction, a path from node 1 to node $n$ corresponds to binary string where the $i$th bit depends on whether $e_i$ or $e'_i$ is taken. Additionally, a path has total probability $k/d$ only if the binary string matches $k$ of the templates. This completes the first reduction. □

Now it remains to show the following:

**Lemma 3.2.** *The string template matching problem is NP-hard.*

*Proof.* We will transform the 3-SAT problem to the string template matching problem. Let the clauses of the problem be denoted by $c_1, \ldots, c_m$. Let the variables be denoted $v_1, \ldots, v_p$. Now, we will create $3m$ templates of length $p + 2m$ each. For each clause $c_i$ we will create 3 templates $s^{(i,1)}, s^{(i,2)}, s^{(i,3)}$. The construction procedure is as follows. Initially, the templates contain all wildcards. Suppose $v_j$ is the first variable occuring in $c_i$. If $v_j$ occurs positively then set $s_j^{(i,1)} = 1$, otherwise set $s_j^{(i,1)} = 0$. For the other two variables in $c_i$, do the same for the other two templates. Finally, set the $p + 2i - 1$ and $p + 2i$ entries of $s^{(i,1)}, s^{(i,2)}, s^{(i,3)}$ to be 00, 01, and 10 respectively (these entries ensure that at most one of the 3 templates can match a given bit string).

Now consider a binary string. We will interpret the first $p$ entries as a truth assignment to $v_1, \ldots, v_p$. We may assume that the ensuing bits are a sequence of 00, 01, and 10s. Then it should be clear that the truth assignment satisfies clause $c_i$ if and only if one of the templates $s^{(i,1)}, s^{(i,2)}, s^{(i,3)}$ matches the string. Therefore, we can conclude that there is a satisfying truth assignment if and only if there is a string matching exactly $m$ templates. □

## 3.2 LOWER BOUNDS ON THE OPTIMAL PATH PROBABILITY

Let $OPT$ denote the probability that the most reliable path has not failed. That is,

$$OPT = \max_{\pi \in \Pi} \Pr[\pi \text{ has not failed}].$$

In this subsection, we give some methods for computing lower bounds on $OPT$. Such methods provide some theoretical insight into the structure of the problem of finding the most reliable path. Additionally, the results presented here may serve as a foundation for the development of future approximation algorithms for solving this problem.

Suppose $\pi$ is a path from $s$ to $t$ in the graph $G$. Let $f(\pi)$ denote the log-probability of the event that $\pi$ has

not failed:

$$f(\pi) = \log\left[\sum_i \Pr[\mathcal{X} = i] \Pr[\pi \text{ has not failed}|\mathcal{X} = i]\right].$$

Now, define

$$g(\pi) = \sum_i \Pr[\mathcal{X} = i] \log \Pr[\pi \text{ has not failed}|\mathcal{X} = i].$$

Because log is concave, by Jensen's inequality we immediately have that

$$g(\pi) \leq f(\pi). \qquad (2)$$

**Proposition 3.3.** *We can compute the optimal value of $\max_\pi g(\pi)$ in polynomial time.*

*Proof.* We can apply a simple dynamic programming algorithm. Suppose for a vertex $v$ we want to find an optimal path $\pi$ under the criterion from $s$ to $v$. Suppose $v$'s ancestors are given by $u_1, \ldots, v_n$ and the optimal paths to those vertices are given by $\pi^{(i)}$. Then note that the optimal path $\pi$ must contain one of the $\pi^{(i)}$ as a prefix. This is because we can rewrite $g(\pi)$ as a purely additive function over edges of the path. Therefore, if the prefix were suboptimal, replacing it with the optimal prefix would only improve the objective. Thus, we can compute the optimal path $\pi$ to $v$ just by considering each of the optimal paths to the ancestors. □

Note that in the case of the original problem of maximizing $f(\pi)$, we do not have this property where the optimal path to an ancestor must be the prefix of the optimal path to the goal vertex. This is because the objective $f(\pi)$, unlike $g(\pi)$, is not purely additive. Now let $\pi^* = \arg\max_\pi f(\pi)$ and let $\sigma^* = \arg\max_\pi g(\pi)$. We note the following:

**Lemma 3.4.** *The following inequalities hold:*

$$g(\pi^*) \leq g(\sigma^*) \leq f(\sigma^*) \leq f(\pi^*).$$

*Proof.* The first inequality holds because $\sigma^*$ maximizes $g$. The second inequality holds because of (2). The last inequality holds because $\pi^*$ maximizes $f$. □

## 4 ALGORITHMS

In this section, we describe some methods for finding the most reliable path. First, we describe a method for finding an exact solution that might take exponential time in the worst case. Afterwards, we describe a method for finding approximate solutions that is more efficient.

### 4.1 EXACT ALGORITHM

We can formulate an integer program which produces an exact solution to the problem of finding the most reliable path. Let the vertices of the graph be denoted $\{1, 2, \ldots, n\}$, where the source vertex is 1 and the sink vertex is $n$. Let $p^{(k)}$ denote the probability that the hidden variable $\mathcal{X}$ takes on the value $k$. Let $c_{ij}^{(k)}$ denote the cost of the edge $ij$ (the log-probability) conditioned on the event that $\mathcal{X} = k$. That is, $c_{ij}^{(k)} = \log \Pr[\text{edge } ij \text{ has not failed}|\mathcal{X} = k]$. Let $x_{ij}$ be a variable that is 1 if there is an edge from vertex $i$ to $j$ and 0 otherwise. It should be clear that the following integer program solves the problem:

$$\begin{aligned}
\text{maximize} \quad & \sum_k p^{(k)} \exp\left(\sum_{ij} c_{ij}^{(k)} x_{ij}\right) \\
\text{subject to} \quad & \sum_j x_{ij} - \sum_k x_{ki} = \begin{cases} 1 &, i = 1 \\ 0 &, i = 2, \ldots, n-1 \\ -1 &, i = n \end{cases} \\
& x_{ij} \in \{0, 1\}.
\end{aligned} \qquad (3)$$

The constraints of the program ensure that the variables $x_{ij}$ define a single path from vertex 1 to $n$. The objective is to maximize the reliability of the path.

Clearly, solving the integer program directly is impractical in the worst case. It is natural to consider a relaxation of the above integer program where the integrality constraints $x_{ij} \in \{0, 1\}$ are relaxed to $0 \leq x_{ij} \leq 1$ so that fractional solutions are allowed. Now, we first note that the objective function of (3) is in fact convex:

**Proposition 4.1.** *The objective function of (3) is convex.*

*Proof.* Observe that each function of the form $\sum_{ij} c_{ij}^{(k)} x_{ij}$ is affine. Composition of a convex function with an affine mapping yields a convex function, so each function of the form $\exp\left(\sum_{ij} c_{ij}^{(k)} x_{ij}\right)$ must itself be convex. Finally, any linear combination of convex functions must be convex, so the objective must be convex. □

We make the following observation which may prove useful in computing exact solutions. It turns out that if we can solve the relaxed version of the integer program, then we can immediately calculate an optimal solution to the integer program:

**Theorem 4.2.** *Any optimal solution to the relaxed program can be converted to an optimal solution to the integer program in polynomial time.*

*Proof.* Suppose we have a solution to the relaxed problem. This fractional solution can be viewed as a "flow"

from vertex 1 to vertex $n$. We can now apply a procedure from [Raghavan and Tompson, 1987] to select paths randomly from this flow. That is, we greedily choose some path that has positive flow from 1 to $n$. Then, we "remove" this path from the graph by subtracting the value of the minimum edge from the flow along the path. We store the removed path in our "set of possible paths" and associate to it a probability equal to the value of the minimum edge. Once the procedure completes, we are left with a set of possible paths with associated probabilities summing to 1.

Next, we randomly select a path from the generated set according to the associated probabilities. Let $X_{ij}$ denote the indicator random variable that is 1 if edge $ij$ appears in the selected path. Note that according to this construction, $\Pr[X_{ij} = 1]$ is exactly the original flow found along the edge $ij$ (i.e., the value of $x_{ij}$ in the optimal solution, $x_{ij}^*$). Now, consider the expected value of the objective function:

$$E\left[\sum_k p^{(k)} \exp\left(\sum_{ij} c_{ij}^{(k)} X_{ij}\right)\right]$$
$$= \sum_k p^{(k)} E\left[\exp\left(\sum_{ij} c_{ij}^{(k)} X_{ij}\right)\right]$$
$$\geq \sum_k p^{(k)} \exp\left(E\left[\sum_{ij} c_{ij}^{(k)} X_{ij}\right]\right)$$
$$= \sum_k p^{(k)} \exp\left(\sum_{ij} c_{ij}^{(k)} x_{ij}^*\right)$$
$$\geq OPT.$$

The first inequality follows from Jensen's inequality and the convexity of the function $\exp(x)$. The last inequality follows from the fact that the relaxed program must have a solution exceeding that of the integer program.

Given that the procedure described above produces a path with expectation at least $OPT$, it immediately follows that every possible path chosen by the procedure must achieve probability $OPT$. □

Unfortunately, *maximization* of a non-linear convex objective (as opposed to minimization) over a compact feasible set cannot be solved efficiently in the general case. This is because solutions to the problem occur on the boundaries and extreme points of the feasible set, and so we cannot simply apply gradient descent methods as in the case of minimization. Nevertheless, convex maximization problems have been studied in operations research literature, and various techniques have been developed which can be readily applied to our formulation [De Loera et al., 2006, Horst et al., 2000]. Many of the practical techniques for solving these sorts of problems rely on heuristics rather than theoretical guarantees. As was shown in a previous section, if we demand exact answers, we cannot hope to do better.

### 4.2 APPROXIMATION ALGORITHM

In this subsection we consider an approximation algorithm for the problem of finding the most reliable path. This approach is inspired by techniques for approximation that utilize discretization and rounding [Sahni, 1977]. Supposing that we restrict the number of total possible path probabilities to some small number, we can solve problems in such cases through exhaustive (pseudo-polynomial dynamic-programming) approaches. For example, (in the extreme case) if all edge probabilities are 0 or 1, then it is easy to construct a simple and efficient algorithm for solving the problem.

#### 4.2.1 A Special Case

Let us first consider the special case where there are $n$ total vertices and only $q$ distinct edge probabilities, and furthermore, the edge log-probabilities are integral with values ranging from $\{0, \ldots, -q+1\}$. That is, for all $e \in E$ and $k \in \mathcal{V}$, we have $c_e^{(k)} \in \{0, \ldots, -q+1\}$ where $c_e^{(k)} = \log \Pr[e \text{ has not failed}|\mathcal{X} = k]$. This case is particularly interesting because there are only $nq$ possible probabilities for paths. We can give an algorithm which takes time exponential in the number of possible values the hidden variable can take $d$, but only polynomial in the size of the graph $n$ and the number of distinct probabilities $q$:

**Proposition 4.3.** *In the case that edge log-probabilities are from the set $\{0, \ldots, -q+1\}$, we can compute the optimal path exactly in time $O\left(n^2 (nq)^d\right)$.*

*Proof.* We will give a dynamic programming algorithm. Let $\Pi(v, a_1, \ldots, a_d)$ be any possible path to vertex $v$ where for each $i$, conditioned on the event that $\mathcal{X} = i$, the path does not fail with probability $a_i$. Note that $\Pi(v, a_1, \ldots, a_d)$ can be computed easily in terms of the $\Pi(u, \cdot, \ldots, \cdot)$ for each $u \in N^-(v)$ (we use $N^-(v)$ to denote the set of vertices such that $u \in N^-(v)$ if and only if there exists a directed edge from $u$ to $v$ in $G$). There are $O\left(n(nq)^d\right)$ entries in the dynamic programming "table" $\Pi$, and computing each entry takes $O(n)$ time. Finding the optimal path is trivial once the table is computed. □

To summarize, our algorithm is the following:

1. Let $\Pi(v, c_1, \ldots, c_d)$ give a backwards edge from $v$ defining (recursively) a path $\Pi$ to $v$ from $s$ such that $\Pi$ has cost (log-probability) $c_i$ if $\mathcal{X} = i$. Initially, all $\Pi(v, c_1, \ldots, c_d) = \emptyset$.

2. Sort the vertices $V - \{s\}$ in topological order.

3. For each vertex $v$ in the topological ordering:

   (a) For each possible set of values $c_1, \ldots, c_d$, if there exists an edge $e \in E$ from $u$ to $v$ such that
   $$\Pi\left(u, c_1 - c_e^{(1)}, \ldots, c_d - c_e^{(d)}\right) \neq \emptyset$$
   set $\Pi(v, c_1, \ldots, c_d) = u$.

4. Return the path from $\Pi(t, \cdot, \ldots, \cdot)$ that is most reliable.

### 4.2.2 The General Case

The above algorithm has a running time that depends strongly on the $q$ parameter which governs the number of distinct possible edge failure probabilities. Instead, we would like to be able to solve problems where $q$ is arbitrarily large—this would allow us to solve problems where the edge failure probabilities are arbitrary rational numbers (through simple scaling). In this subsection, we present an approximation algorithm with a running time that does not depend on $q$ that therefore addresses this issue.

Our strategy will be to "coarsen" the granularity of problem so that we can apply the aforementioned dynamic programming solution to a problem with a smaller $q$. Let $\epsilon > 0$ be an error parameter given to us. Then let
$$k = \frac{-\log(1 - \epsilon)}{n}.$$

What we will then do is for each log-probability $\log a$, we will replace it with a new log probability
$$\log a' = \left\lfloor \frac{\log a}{k} \right\rfloor k.$$

Then, we will solve the problem using the previous dynamic programming approach. Let us quantify the error that is introduced by this method:

**Proposition 4.4.** *The coarsening approach gives a $1 - \epsilon$ approximation algorithm that runs in $O\left(n^2 \left(n \left\lfloor \frac{-qn}{\log(1-\epsilon)} \right\rfloor\right)^d\right)$ time.*

*Proof.* All the log-probabilities are non-positive. Therefore, we have that
$$\frac{\log a}{k} - 1 \leq \left\lfloor \frac{\log a}{k} \right\rfloor \leq \frac{\log a}{k},$$
which is equivalent to
$$\log a + \frac{\log(1 - \epsilon)}{n} \leq \log a' \leq \log a$$
by the definition $a'$ and $k$. Exponentiating each side yields a statement about probabilities rather than log-probabilities:
$$a(1 - \epsilon)^{1/n} \leq a' \leq a.$$

So we see that the error introduced to each edge is a factor of $(1 - \epsilon)^{1/n}$. Note however that every path is of length at most $n$, so the maximum error introduced to the probability of a path being reliable is $(1 - \epsilon)^{(1/n) \cdot n}$. Finally note that the maximum error remains the same even when we take the convex combination of multiple paths with the same error. $\square$

Unfortunately, the above is quite unsatisfactory because the running time still depends on $q$. If we make one additional assumption—that is, that the best path has probability of being reliable at most $e^{-q}$, then we can improve the algorithm. This assumption means that the probability of the best path being reliable must be at most the probability of the most unreliable edge failing. We will show that we can dispatch this assumption later. We will use the following trick:

**Lemma 4.5.** *Suppose $x, y, \epsilon, \delta > 0$. The additive inequality $y \leq x + \delta$ implies the multiplicative inequality $y \leq x(1 + \epsilon)$ if $x \geq \delta/\epsilon$.*

*Proof.* In order for the implication to hold, we require that $x(1 + \epsilon) \geq x + \delta$ (the multiplicative inequality must be weaker than the additive one). However, note that this inequality is equivalent to the inequality $x \geq \delta/\epsilon$. $\square$

Let us set $k = \epsilon q/n$. Following the argument from before, if $\pi$ corresponds to the best path, and $\pi'$ corresponds to the best path in the coarsened problem, we have
$$-\log p(\pi) \leq -\log p(\pi') \leq -\log p(\pi) + \epsilon q.$$

However, we have that $-\log p(\pi) \geq q$, so the additive inequality can be converted into a multiplicative one:
$$-\log p(\pi) \leq -\log p(\pi') \leq -\log p(\pi)(1 + \epsilon).$$

This gives us the bound
$$p(\pi') \geq p(\pi)^{1+\epsilon}. \qquad (4)$$

The running time then becomes $O\left(n^2 \left(n \left\lfloor \frac{n^2}{\epsilon} \right\rfloor\right)^d\right)$.

One issue is that we had to assume that the best path has probability (of being reliable) at most $e^{-q}$ in order for us to attain the above performance guarantee. However, we can overcome this assumption by a simple pruning procedure. There are $d\binom{n}{2}$ total possible unique edge probabilities of the form $\Pr[e \text{ has not failed}|\mathcal{X} = k]$ for some $e \in E$ and $k \in \mathcal{V}$. Let these edge probabilities be given in sorted order by $a_1 \leq a_2 \leq \ldots \leq a_{d\binom{n}{2}}$. For each $i$ such that $1 \leq i \leq d\binom{n}{2}$, we run the above approximation algorithm on the graphs with all edges of probability $< a_i$ deleted, producing a path $\pi_i$. Now, we claim that for some choice of $i$, none of the edges from the original optimal path are deleted and the assumption that the best path has probability of being reliable at most $e^{-q}$ is fulfilled. Indeed, this must be the case when $a_i$ is the probability of the lowest probability edge on the original optimal path. Thus, at least one of the $\pi_i$ must meet the approximation guarantee (4). Therefore, we simply choose our path to be the best path out of the $\pi_i$'s

$$\arg\max_{\pi_i} \{p(\pi_i)\}.$$

This extra pruning procedure requires that our algorithm runs for an additional $O(dn^2)$ iterations, so the algorithm still takes only polynomial time in the size of the graph.

We summarize these results in the following theorem:

**Theorem 4.6.** *For any $0 < \epsilon < 1$, there is an approximation algorithm that finds a path with reliability $OPT^{1+\epsilon}$ in time polynomial in the size of the graph and $1/\epsilon$ and exponential in $d$, where $OPT$ is the probability that the optimal path does not fail and $d$ is the number of values the hidden variable can take.*

### 4.3 MULTIPLE HIDDEN VARIABLES AND ADDITIONAL STRUCTURE

In this paper we have considered only the simple case where there is a single hidden variable influencing the edge failure variables of a network. We believe a natural extension to this problem that is worthy of future investigation is the case where there may be multiple hidden variables—or even a general graphical model structure—influencing the edge failure variables. Clearly any setting with multiple hidden variables can be converted to a setting where there is a single vector-valued hidden variable, but performing this conversion discards combinatorial structure in the problem which may be exploitable for purposes of efficiency.

For example, one could consider a case where there are multiple hidden variables $\mathcal{X}_1, \ldots, \mathcal{X}_k$ and where each hidden variable $\mathcal{X}_i$ affects only a restricted subset $S_i \subseteq E$ of the edges of the graph. It is not difficult to extend the algorithms that we have given in this paper to such a case and show that the multiple-variable problem can often be solved more efficiently than the equivalent single-variable problem by taking advantage of the structure of the problem. We suspect that richer combinatorial structures such as general graphical models can be exploited in similar manners as well.

## 5 CONCLUSION

In this paper we have presented a new inference problem that introduces the notion of hidden state to traditional graph reachability problems. We believe algorithms that solve this problem will prove useful in numerous applications involving finding minimum-cost paths or most-likely sequences of states, including network QoS routing, pattern recognition, and parsing. We have given theoretical results which help to characterize the problem, as well as algorithms (exact and approximate) for tackling the problem. We have also briefly discussed directions of future research involving more structured variants of the problem that we believe will be fruitful for further investigation.